\RequirePackage{fix-cm}
\documentclass[smallextended]{svjour3}       
\smartqed  
\usepackage{graphicx}
\usepackage{subfigure}
\usepackage{epsfig}
\usepackage{amsmath}
\usepackage{amsfonts}   
\usepackage{amssymb}
\usepackage{amsmath}
\usepackage{amsfonts}   
\usepackage{amssymb}

\def\e{{\epsilon}}

\def\norm#1{\|#1\|}

\renewcommand{\t}{\tau}

\newcommand{\mlabel}[1]{\label{#1}
}

\newcommand{\pf}{ \par \vspace{1ex} \noindent {\sc Proof} \hspace{2mm}}
\newcommand{\epf}{$ \quad \Box$ \par \vspace{1ex}}

\newcommand{\vect}[1]{\mathbf{#1}}
\providecommand{\abs}[1]{\lvert#1\rvert}
\providecommand{\norm}[1]{\lVert#1\rVert}

\begin{document}

\title{Spreading speeds and traveling waves for non-cooperative integro-difference systems\thanks{This manuscript is available at arXiv:1003.1600}}


\titlerunning{Minimum speeds and traveling waves}        

\author{Haiyan Wang  \and  Carlos Castillo-Chavez
}

\authorrunning{Wang, Castillo-Chavez} 

\institute{Haiyan Wang \at
           Division of Mathematical and Natural Sciences\\
           Arizona State University\\
           Phoenix, AZ 85069, USA\\
           \email{wangh@asu.edu}
           \and
           Carlos Castillo-Chavez \at
           Mathematics, Computational and Modeling Sciences Center\\
           Arizona State University\\
           PO Box 871904, Tempe, AZ 85287, USA\\
           School of Human Evolution and Social Change\\
           Arizona State University\\
           Tempe, AZ 85282, USA\\
           Santa Fe Institute \\
           1399 Hyde Park Road,  Santa Fe, NM 87501, USA\\
           Biological Statistics and Computational Biology\\
           Cornell University\\
           Ithaca, NY 14853 - 2601\\
           \email{ccchavez@asu.edu}
}

\date{Received: date / Accepted: date}

\maketitle

\begin{abstract}
The study of spatially explicit integro-difference systems when the local population dynamics are given in terms of discrete-time generations models, has gained considerable attention over the past two decades. These nonlinear systems arise naturally in the study of the spatial dispersal of organisms. The brunt of the {\it mathematical} research on these systems, particularly, when dealing with  {\it cooperative} systems, has focused on the study of the existence of  traveling wave solutions and the characterization of their spreading speed. Here, we characterize the minimum propagation (spreading)  speed, via the convergence of initial data to wave solutions, for a large class of {\it non cooperative} nonlinear systems of integro-difference equations. The spreading speed turns out to be  the one with slowest speed from a family of non-constant traveling wave solutions. The applicability of these theoretical results is illustrated through the explicit study of an integro-difference system with {\em local} population dynamics governed by Hassell and Comins'  non-cooperative competition model (1976).    The corresponding integro-difference nonlinear systems that results from the redistribution of individuals via a dispersal kernel is shown to satisfy  conditions that guarantee the existence of minimum speeds and traveling waves.  This article is dedicated to Simon A. Levin whose contributions to the fields of ecology, evolutionary biology, and the environmental sciences have driven and inspired the research of generations of mathematicians, mathematical biologists, and life and social scientists, around the world for over four decades.
\end{abstract}

\section{Introduction}\label{induc}

Finding and developing macroscopic descriptions for the dynamics and  behavior of heterogeneous large ensembles of individuals subject to
ecological forces like dispersal continues to provide challenges and opportunities for mathematical and biological scientists. Over the past century,
 particular attention has been placed on the study of  the role played by dispersal  in shaping plant communities, in helping understand  biological invasions, in assisting in the quantification and control of the spread of infectious disease, or in disentangling the dynamics of marine open-ocean and intertidal systems, to name but a few examples. The work of pioneers like  Aronson  \cite{Aronson1975,Aronson1978}, Fisher  \cite{Fisher}, Hadeler  \cite{Hadeler1975,Hadeler1988,Hadeler1999}, Kolmogorov \cite{Kolmogorov}, Levin \cite{Levin}, Okubo \cite{Okubo}, Skellam \cite{Skellam}, Slobodkin \cite{Kierstad},  Weinberger \cite{Weinberger1982} and the subsequent cadre of {\em distinguished}  mathematicians and theoreticians across the world who have worked at this interface, set not only the foundation of an important and  fertile area of interdisciplinary research (ecology, mathematics, and evolutionary biology) but in the process it has inspired novel mathematical research while being re-energized by unsolved questions in emerging  fields like urban ecology and sustainability and the challenges and opportunities posed by the growing body of  research on the co-evolving dynamics of socio-biological systems \cite{Crow2010} \cite{Levin2009}.

The {\em theme} of finding {\it mathematical} macroscopic descriptions for the spatial dynamics of heterogeneous  large-ensembles of populations was set in ``motion" by  the fundamental ecological contributions of Skellam (1951) \cite{Skellam},  Kierstad and Slobodkin (1953) \cite{Kierstad}, Levin and Paine (1974) \cite{Levin}, Okubo (1980)  \cite{Okubo}, and others. The study of integro-difference equations dispersal models in the mathematical literature has its origins  in the study of the coupled  spatial dynamics of organisms with discrete {\it primarily} non-overlapping (but see \cite{Rios-Soto}) local dynamics with dispersal processes modeled via re-distribution kernels \cite{Aronson1975,Aronson1978}.

Early models for the dispersal of invasive species used nonlinear reaction-diffusion equations, with the prototype provided by Fisher's Equation \cite{Fisher}. The primary motivation or emphasis have always been on the characterization of the speed of propagation of species invading unoccupied habitats. The seminal contributions of Fisher \cite{Fisher} and Kolmogorov, Petrowski, and Piscounov \cite{Kolmogorov}, Aronson and Weinberger \cite{Aronson1975,Aronson1978}  jointly handled the mathematical challenges posed by their efforts to classify rigorously the concept of {\it speed of  propagation} for quite general continuous in space and discrete time, integro-difference equations. Weinberger \cite{Weinberger1982} and Lui \cite{Lui1989} research expanded the mathematical foundation for the theory of spreading speeds and traveling waves, through their analysis  of traveling waves via the convergence of initial data to wave solutions, in the context of {\em cooperative} operators.   Recently, Weinberger, Lewis and  Li  made additional contributions \cite{Weinberger2002-1,Weinberger2002,Weinberger2005,Weinberger2007}. The mathematical analyses of integro-difference spatially explicit systems enhances the understanding of the dynamics of introduced species like weeds or pests in terrestrial systems, or the study of the impact of dominant alien species in freshwaters while generating additional challenges and opportunities to mathematicians, whose interests, are driven by the study of challenging dynamical systems.

The pervasiveness  of overcompensation in biological systems implies that integro-difference equations models are in general {\it non-cooperative}, and therefore, existing theoretical work has yet to address effectively the mathematical consequences of  {\it non-cooperative} local dynamics on dispersal. In other words, the incorporation of biological forces/mechanisms that drive population overcompensation leads to mathematical models whose dynamics have yet to be satisfactorily teased out in the context of relevant biological settings.  Deep mathematical challenges remain  \cite{Kot1992}.  The research in this manuscript does not start in the vacuum since relevant mathematical work for non-cooperative systems has been carried out by several researchers. Thieme \cite{Thieme1979} showed, in the context of a general model with non-monotone growth functions, that the asymptotic spreading speed  could still be obtained with the aid of carefully constructed monotone functions. Hsu and Zhao \cite{hsu2008} and Li, Lewis and Weinberger \cite{LiLewis2009} just extended the theory of spreading speeds in the context of non-monotone integro-difference equations. Their extensions relied on two methods: the construction of two monotone operators (with appropriate properties) and the application of fixed point theorems in Banach spaces--an approach also used in Ma \cite{ma2007} and Wang \cite{Hwang2009} to establish the existence of traveling wave solutions of reaction-diffusion equations. The results in this manuscript on the speed of propagation for non-cooperative systems in
the context of integro-difference equations rely on the spreading results for monotone systems in Weinberger et al. \cite{Weinberger2002-1}.

We highlight our results in the context of  a two-dimensional nonlinear discrete system describing the local nonlinear dynamics of  two competing  species with discrete reproduction cycles \cite{Hassell}.  The model focuses on the growth and spread of these competing species with their population densities at generation $n$ and spatial location $x$ being tracked by the state variables $X_n(x)$ and $Y_n(x)$, respectively. The system is a natural extension of the classical single population ``scramble" competition model of Ricker \cite{Brauer2001}. Specifically, the non-spatial interference-competition model of Hassell and Comins is given by the following system of  coupled nonlinear difference equations:
\begin{equation}\label{eqoper011}
\begin{split}
X_{n+1}(x)&=X_{n}(x)e^{r_1-X_n(x)-\sigma_1 Y_n(x)}\\
Y_{n+1}(x)&=Y_n(x)e^{r_2-Y_n(x)-\sigma_2 X_n(x)}
\end{split}
\end{equation}
where $r_1,r_2, \sigma_1, \sigma_2$ are all positive constants.

The possibility that individuals in the above two populations may disperse to different sites is modeled with a redistribution kernel $k_i(y)$. Hence, a discrete-time model, where individuals interact locally according to Model (\ref{eqoper011}),  can be naturally formulated via a system of coupled nonlinear integro-difference equations. Hence, we have that
\begin{equation}\label{eqoper01}
\begin{split}
X_{n+1}(x)&=\int_{\mathbb{R}}k_1(x-y)X_{n}(y)e^{r_1-X_n(y)-\sigma_1 Y_n(y)}dy\\
Y_{n+1}(x)&=\int_{\mathbb{R}}k_2(x-y)Y_n(y)e^{r_2-Y_n(y)-\sigma_2 X_n(y)}dy
\end{split}
\end{equation}

\noindent where the dispersal of the i-species is modeled by a redistribution kernel $k_i, i = 1, 2$ that depends just on the signed distance
$x-y$, connecting the ``birth" $y$ location  and the ``settlement" location $x$. In other words, $k_i(y)$ is a homogenous ``probability" kernel that satisfies $\int_{-\infty}^{\infty}k_i(y)dy=1$.

 Since the above system is {\em non-cooperative} in general, it is in such a  context  that new results will be formulated, and illustrated but first, we introduce the notation that will be used for the explicit mathematical formulation of the dynamics of two-interacting, {\em dispersing}, and competing populations.
Consequently, $\beta, \beta^{\pm}, F, F^{\pm}, r, u,v$ are used to denote vectors in $\mathbb{R}^N$ or $N$-vector valued functions while  $x,y,\xi$ are used to denote variables in $\mathbb{R}$. The use of  $u=(u^i)$ and $v=(v^i) \in \mathbb{R}^N$ allow us to define $u \geq v$ whenever $u^i \geq v^i $ for all $i$; and
$u \gg v$ whenever $u^i > v^i$ for all $i$. We further define for any $r=(r^i) \gg 0, r \in \mathbb{R}^N$  the ${R}^N$-interval
$$
[0,r]= \{ u: 0 \leq u \leq r, u \in \mathbb{R}^N\}\subseteq \mathbb{R}^N
$$
and
$$
\mathcal{C}_{r}= \{u=(u^1,...,u^N): u^i \in C(\mathbb{R}, \mathbb{R}), 0\leq u^i(x) \leq r^i,  x\in \mathbb{R},\; i=1,...,N\},
$$
where $C(\mathbb{R}, \mathbb{R})$ is the set of all continuous functions from $\mathbb{R}$ to $\mathbb{R}$. Our focus will be on the set  $\mathcal{C}_{\beta^{+}}, \beta^{+} \gg 0$.

Specifically, we consider the system of integro-difference equations
\begin{equation}\label{eq1}
u_{n+1}=\mathcal{Q}[F(u_{n})];
\end{equation}
where
 $u_n=(u^i_n) \in \mathcal{C}_{\beta^{+}}$, $F(u)=(f_i(u));$
$$\mathcal{Q}[F(u)]=(\mathcal{Q}^i[F(u)]);$$
\begin{equation*}
\begin{split}
\mathcal{Q}^i[F(u)](x)&=\int_{\mathbb{R}}k_i(x-y)f_i\big(u(y)\big)dy;\\
\end{split}
\end{equation*}
$u_n(x)$ is the density of individuals at point $x$ and time/generation
$n$; $F(u)$ is the density-dependent fecundity (local growth rate); and $k_i(x-y)$ (dispersal kernel)
models the dispersal of $u$,  assumed to depend only on the signed distance
$x-y$ between the location of ``birth" $y$ and the ``settlement" or ``landing'' location
$x$. As noted before, $k_i(x-y)$ can be viewed as a probability kernel since $\int_{-\infty}^{\infty}k_i(x)dx=1$.
The notation $\mathcal{Q}[F(u_{n})]$ is slightly different from these used in \cite{hsu2008,LiLewis2009,Lui1989,Weinberger2002-1} and hence those wishing to compare results must account for this. Specifically, no $F$ can be found in the standard literature notation.  Here,  $F$ has been included to
carry out the proofs involving non-monotone systems effectively.

The integro-difference system (\ref{eq1}) models the reproduction and
dispersal of a time-synchronized species where all individuals
first undergo reproduction, then redistribute their offspring,  and then proceed to
reproduce again. The goal is to carry out the characterization of the spreading speed in a system involving a rather general non-cooperative system (\ref{eq1}) as the slowest speed of a family of non-constant traveling wave solutions of (\ref{eq1}).

\section{Non-cooperative Systems' Results}\label{rest}
The focus  is on the characterization of the speeds of propagation for (\ref{eq1}) when the system is non-cooperative. As it is typical in mathematics, we make use of prior results established results for {\em cooperative} systems (\cite{Weinberger2002-1}).
The existence of two additional monotone operators $F^{\pm}$ with the properties that the first lies above  $F$  and the second below $F$ is required by our method of proof. The use of this approach is motivated by the work on non-monotone equations carried out in \cite{Thieme1979,hsu2008,LiLewis2009,ma2007,WeinbergerKS2009,Hwang2009}. Specifically, we observe that $F^{\pm}$ can be ``constructed" via piecewise functions made up of ``pieces' of $F$ and
the incorporation of appropriate constants. If $F$ happens to be monotone, then  $F^{\pm}=F$. We introduce additional technical assumptions below. The assumptions are critical because the feasibility of the mathematical analysis depends on whether or not the components of our problem meet them:

\begin{enumerate}
  \item[(H1)] For $i=1,...,N$, $k_i(\t)\geq 0$ is integrable on $\mathbb{R}$,  $k_i(\t)=k_i(-\t), \t \in \mathbb{R},$
and $\int_{\mathbb{R}}k_i(\t)d\t =1, \int_{\mathbb{R}}k_i(\t)e^{\lambda \t}d\t< +\infty,$
for all $\lambda>0.$
  \item[(H2)]
  \begin{itemize}
             \item[(i)] Given that $F: [0,\beta^{+}] \to [0,\beta^{+}]$ is a continuous, twice piecewise continuous differentiable function with
              $$ 0\ll \beta^{-}=(\beta^-_i)\leq \beta=(\beta_i) \leq \beta^{+}=(\beta^+_i),$$ it is assumed that there exist continuous, twice piecewise continuous differentiable functions
              $F^{\pm}=(f^{\pm}_i): [0,\beta^+] \to [0,\beta^{+}]$ such that
              for $u\in [0,\beta^+]$, $$F^{-}(u) \leq F(u) \leq F^{+}(u).$$
              \item[(ii)] $F(0)=0, F(\beta)=\beta$ and there is no other positive equilibrium of $\mathcal{Q}[F]$ between $0$ and $\beta$
               (that is, there is no constant $v \neq \beta$ such that $F(v)=v, 0 \ll v \leq \beta$).
              $F^{\pm}(0)=0, F^{\pm}(\beta^{\pm})=\beta^{\pm}$ and there is no other positive equilibrium of $\mathcal{Q}[F^{\pm}]$
              between $0$ and $\beta^{\pm}$. $F$ has a finite number of equilibria in $[0, \beta^+]$.
              \item[(iii)]
              $F^{\pm}$ are nondecreasing functions on $[0, \beta^+]$ and $F^{\pm}(u)$ and $F(u)$ have the same Jacobian at $0$.
              \end{itemize}
\end{enumerate}

Assumptions (H1-H2) do not suffice if the goal is to characterize the speeds of propagation for (\ref{eq1}).
The assumption (H3), which includes the requirement that the operator grows less than its linearization along the particular function $\nu_{\mu} e^{-\mu x }$, is essential and implies that the operator $\mathcal{Q}$ does not display an Allee effect for this particular function (see \cite{Weinberger2002-1}). Assumption (H3), explicitly formulated below, is satisfied by several {\em biological systems} of interest this will be highlighted in our example.  Assumption (H3), therefore does not severely handicap the usefulness of the results in this manuscript.

The need for Frobenius' theorem stating that any nonzero irreducible matrix with nonnegative entries has a unique principal positive eigenvalue with a corresponding principal eigenvector ``made up" of strictly positive coordinates is implicit in  Assumption (H3). The formulation of (H3) depends on the concept of irreducibility. A matrix is irreducible if it is not similar to a lower triangular matrix with two blocks via a permutation (See \cite{Horn,Weinberger2002-1}). By reordering the coordinates, one can put any matrix into a block lower triangular form, then we say that the matrix is in {\it Frobenius form} if all the diagonal blocks are irreducible (an irreducible matrix consists of the single diagonal block which is the matrix itself). Here we use the definition of {\it Frobenius form} in Weinberger et al. \cite{Weinberger2002-1}.
Following the approach in \cite{Weinberger2002-1}, that for each $\mu>0$,  the $N\times N$ matrix $B_{\mu}$ that results from the linearization of (\ref{eq1}) at $0$, namely
\begin{equation}\label{matrix1}
B_{\mu}=(b^{i,j}_{\mu})=\big(\partial_j f_i(0) \int_{\mathbb{R}}k_i(s)e^{\mu s}ds\big),
\end{equation}
where $b^{i,j}_{\mu}$ is the $(i,j)$ entry of the matrix, is  in {\it Frobenius form} (\cite{Weinberger2002-1}).
In other words, it is assume that the required reordering has been
done for $B_{\mu}$ (\cite{Weinberger2002-1}).
If we now let $\lambda(\mu)$ denote the principal eigenvalue of the first diagonal blocks, we reach the formulation required by Assumption (H3):

\begin{enumerate}
  \item[(H3) ] \begin{itemize}
   \item[(i)] Assume that $B_{\mu}$ is in Frobenius form and that the principal eigenvalue, $\lambda(\mu)$, of the first diagonal block is strictly larger than
   the principal eigenvalues of other diagonal blocks. Further, let's assume that  $B_{\mu}$  has a positive eigenvector $ \nu_{\mu}=(\nu^i_{\mu})\gg 0$ corresponding to $\lambda(\mu)$ with the additional requirement that $\lambda(0)>1.$
   \item[(ii)] For each $\mu>0$ and $\alpha >0$, we let $v^{\pm}=(v^{\pm}_i)=(\min\{\beta^{\pm}_i, \nu^i_{\mu} \alpha\}),$ and assume that $$
   F^{\pm}(v^{\pm}) \leq B_{0}v^{\pm}.
   $$

   \item[(iii)] For every sufficiently larger positive integer $k$, there is a small constant vector $\omega=(\omega^i)\gg 0$ such that
$$
F^{\pm}(u)\geq (1-\frac{1}{k})B_{0} u, \;\; u \in [0, \omega],
$$
  \end{itemize}
\end{enumerate}

\noindent It follows from (H1) that  $\lambda(\mu)$ is an even function. In fact, it was shown by Lui \cite{Lui1989} that $ \ln \lambda(\mu)$
is a convex function and therefore, $ \ln \lambda(\mu)$ achieves its minimum at $\mu=0$ and, therefore the assumption that $\lambda(0)>1$ implies that $ \ln \lambda(\mu) > 0.$ The statement in Proposition \ref{lmeigen-1}  below which is critical to the rest of analysis that leads to the main result and it involves the following function of the largest principal eigenvalue $\lambda(\mu)$$$
\Phi(\mu)=\frac{1}{\mu} \ln \lambda(\mu) > 0.
$$
Part (5) of Proposition \ref{lmeigen-1} highlights the use of this function in the construction of  lower solutions and estimates of the traveling wave solutions.

\begin{proposition}\mlabel{lmeigen-1}  Assume that $(H1)-(H3)$ hold. Then
\begin{enumerate}
  \item [1] $\Phi(\mu) \to \infty$ as $\mu \to 0;$
  \item [2] $\Phi(\mu)$ is decreasing as $\mu=0^+;$
  \item [3] $\Phi'(\mu) $ changes sign at most once on $(0, \infty)$
  \item [4] $\Phi(\mu)$ has a  minimum $c^*>0$.
  \item [5] For each $c > c^*$, there exist $\Lambda_{c}>0$ and $\gamma \in (1,2)$ such that
$$\Phi(\Lambda_{c})=c,\;\; \Phi(\gamma \Lambda_{c})<c.$$
\end{enumerate}
\end{proposition}

Parts (1)-(4) of Proposition \ref{lmeigen-1} are essentially due to Lui \cite{Lui1989}. However, Lui's results only guarantee that $c^* \geq 0.$
The proof  of the strict inequality, that is, that $c^*>0$, is found in the Appendix.  Since $\lambda(\mu)$ is a simple
root of the characteristic equation of an irreducible block, it can be shown that $\lambda(\mu)$ is twice continuously differentiable on $\mathbb{R}$.
Part (5) is a direct consequence of the results stated in Parts (1)-(4).

A traveling wave solution $u_n$ of (\ref{eq1}) is defined as a solution of the form  $u_n(\xi)=u(\xi-cn), u \in C(\mathbb{R}, \mathbb{R}^N)$.
The theorems that guarantee the existence of traveling wave solutions for cooperative systems have been already established (e.g. \cite{Weinberger2005}).
 It also has been established that the asymptotic spreading speed, for such systems, can be characterized as the speed of the slowest non-constant traveling wave
solution for  monotone operators \cite{Weinberger2005} and for scalar equations \cite{Weinberger1978,Weinberger1982,hsu2008,LiLewis2009}.

 We start with the statement of Theorem \ref{th30}, the main theorem, which  {\it generalizes} results previously established for cooperative systems
 to non-cooperative systems. Some parts of Theorem \ref{th30} such as  the asymptotic behavior of  traveling waves are new even for cooperative systems.
 The new information about cooperative systems are, in fact, required to be able to carry out the proofs of the results for non-cooperative systems.
 The details associated with the proof of the {\em main} result are collected in a series of lemmas and theorems all collected in the following sections.

The two major new contributions in this paper are:  1) for a large class of non-monotone systems (\ref{eq1}),
 the question of the existence of the  minimum speed of propagation  is settled (Theorem \ref{th30}(i-ii)) and this speed is characterize as the speed of the slowest non-constant traveling wave solution (Theorem \ref{th30}(iii-v)); and  2) in the case of {\it competition} model, a direct application of the main theorem  helps identify simple and meaningful conditions needed for the existence of traveling waves with the {\em minimum} speed of propagation
(Theorem \ref{th33}). That is, what is required to guarantee  the success of a biological invasion. It is worth re-iterating that the application of the results in this manuscript to the study of relevant monotone operators case \cite{Weinberger2005,Weinberger1982} does give additional information. In fact, these results help
explicitly characterize the asymptotic behavior of traveling waves via the careful analysis of eigenvalues and upper-lower solutions. This analysis was not done before
\cite{Weinberger2005,Weinberger1982} most likely because the focus was exclusively in establishing the existence of traveling waves. The
results and analysis for the $n$-dimensional case is typically harder. Our approach works because the analysis of the $n$ dimensional case is closely related to  structure of the eigenvalues and corresponding eigenvectors, an analysis that is embedded in our study of the relevant monotone operators.

The following theorem summarizes the main results.
\begin{theorem}\label{th30} Assume $(H1)-(H3)$ hold.  Then the following statements are valid:
\begin{enumerate}
\item [(i)] For any $u_0 \in \mathcal{C}_{\beta}$ with compact support and $0 \leq u_0 \ll \beta$, the solution $u_n$ of (\ref{eq1}) satisfies
$$
\lim_{n \to \infty}  \sup_{\abs{x}\geq nc} u_n(x) = 0, \text{ for } c> c^*
$$

\item [(ii)] For any strictly positive vector $\omega \in \mathbb{R}^N$, there is a positive $R_{\omega}$ with the property
 that if $u_0 \in \mathcal{C}_{\beta}$ and $u_0 \geq \omega $ on an interval of length $2R_{\omega}$, then the solution $u_n(x)$ of (\ref{eq1})
 satisfies
$$
\beta^- \leq \liminf_{n \to \infty} \inf_{\abs{x}\leq nc} u_n(x) \leq  \beta^+, \text{ for } 0< c< c^*
$$

\item [(iii)] For each $c > c^*$ (\ref{eq1}) admits a traveling wave solution $u(\xi-cn)=(u^i(\xi-cn))$ such that
$0 \ll u(\xi) \leq \beta^+, \xi \in \mathbb{R}$,
$$\beta^-\leq \liminf_{\xi \to -\infty}u(\xi) \leq \limsup_{\xi \to -\infty}u(\xi)\leq \beta^+$$
$\lim_{\xi \to \infty}u(\xi)=0$ and
\begin{equation}\label{aystomth112}
\lim_{\xi \to \infty} u(\xi)e^{\Lambda_{c} \xi}=\nu_{\Lambda_{c}}.
\end{equation}
If, in addition, $F$ is non-decreasing on $\mathcal{C}_{\beta}$, then $u$ is non-increasing on $\mathbb{R}$.

\item [(iv)] For $c = c^*$ (\ref{eq1}) admits a non-constant traveling wave solution $u(\xi-cn)=(u^i(\xi-cn))$ such that
$0 \leq  u(\xi) \leq \beta^+, \xi \in \mathbb{R}.$

\item [(v)] For $0<c<c^*$ (\ref{eq1}) does not admit a traveling wave solution $u_n(\xi)=u(\xi-cn)$  such that
$u \in \mathcal{C}_{\beta^+} $ with $ \liminf_{ \xi \to -\infty} u(\xi) \gg 0$ and $u(+\infty)=0.$
\end{enumerate}
\end{theorem}

\begin{remark}\label{rem1}
When $F$ is monotone, $F^{\pm}=F, \beta^{\pm}=\beta$.
\end{remark}

\begin{remark}\label{rem200}
The assumption that $F$ has a finite number of equilibria in $[0, \beta^+]$ is only used in the proof of Theorem \ref{th30} (iv), and 
can be further relaxed. In fact, as long as for some component $i$ and a sufficiently small positive number $\delta$, 
$u=(u^i) \geq 0$ with $u^i=\delta$ are not equilibria of $F$, the conclusion is still valid from the proof.  
\end{remark}

We shall establish Theorem \ref{th30} in Sections \ref{spreadings} and \ref{proofMono}.

\section{Spreading Speeds}\label{spreadings}

Our results on the speed of propagation for non-cooperative systems  make use of Theorem \ref{th20} below which collects
the properties of the spreading speed $c^*$ for monotone systems as established in Weinberger, Lewis and  Li \cite{Weinberger2002-1}.
Theorem \ref{th20} extends the related spreading results in Lui \cite{Lui1989} to systems of monotone recursive operators with more than two equilibria.
The operator at the center of this manuscript may support more than two equilibria with one lying at the boundary as in \cite{Weinberger2002-1} (see Section \ref{example}).

\begin{theorem} \label{th20}(Weinberger, Lewis and  Li \cite{Weinberger2002-1} [Lemma 2.2, Theorem 3.1]) Assume $(H1)-(H3)$ hold. Further assume that $f^i(x), i=1,...,N$ is \textbf{non-decreasing}.  Then the following statements are valid:
\begin{enumerate}
\item [(i)] For any $u_0 \in \mathcal{C}_{\beta}$ with compact support and $0 \leq u_0 \ll \beta$, the solution $u_n(x)$ of (\ref{eq1}) satisfies
$$
\lim_{n \to \infty}  \sup_{\abs{x}\geq nc} u_n(x) = 0, \text{ for } c> c^*
$$

\item [(ii)] For any strictly positive vector $\omega \in \mathbb{R}^N$, there is a positive $R_{\omega}$ with the property
 that if $u_0 \in \mathcal{C}_{\beta}$ and $u_0 \geq \omega $ on an interval of length $2R_{\omega}$, then the solution $u_n(x)$ of (\ref{eq1})
 satisfies
$$
\liminf_{n \to \infty} \inf_{\abs{x}  \leq nc} u_n(x)= \beta, \text{ for } 0< c< c^*
$$
\end{enumerate}
\end{theorem}

It is clear that $\mathcal{Q}[F^{\pm}]$ are monotone (order preserving) on $\mathcal{C}_{\beta^+}.$ That is, if $u,v \in \mathcal{C}_{\beta^+} $
and $u(x) \leq v(x), x \in \mathbb{R}$, then
$$
\mathcal{Q}[F^{\pm}(u)](x) \leq \mathcal{Q}[F^{\pm}(v)](x),\;x \in \mathbb{R}.
$$
Further, for $u=(u^i) \in \mathcal{C}_{\beta^+}$ and $x \in \mathbb{R}$, we have
\begin{equation*}
\begin{split}
f_i^{-}\big(u(x)\big) \leq f_i\big(u(x)\big) \leq f_i^{+}\big(u(x)\big), i=1,...,N.
\end{split}
\end{equation*}
and therefore
\begin{equation*}
\begin{split}
\mathcal{Q}[F^{-}(u)](x) \leq  \mathcal{Q}[F(u)](x) \leq \mathcal{Q}[F^{+}(u)](x),\; x \in \mathbb{R}.
\end{split}
\end{equation*}

We are now able to establish Part (i) and (ii) of Theorem \ref{th30} by following essentially the proof for the scalar cases found in \cite{hsu2008,LiLewis2009}.

\textbf{Proof} of Parts (i) and (ii) of Theorem \ref{th30}.

Part (i).  For a given $u_0 \in \mathcal{C}_{\beta}$ with compact support, let $u_n$ be
the $n$-th iteration of $\mathcal{Q}[F]$ starting from $u_0$ and let $u^{+}_n$ be
the $n$-th iteration of $\mathcal{Q}[F^+]$ starting from $u_0$. By (H2), we have
$$
0 \leq u_n(x) \leq u_n^{+}(x), x \in \mathbb{R}, n >0.
$$
Thus for any $c>c^*$, it follows from Theorem \ref{th20} (i) that $$
\lim_{n \to \infty} \sup_{\abs{x}\geq nc} u^{+}_n(x) = 0,
$$
and hence
$$
\lim_{n \to \infty} \sup_{\abs{x}\geq nc} \abs{u_n(x)} = 0,
$$
Part (ii). Let $u_n,u^{+}_n$ be the $n$-th iteration of $\mathcal{Q}[F],\mathcal{Q}[F^+] $ starting from $u_0$ respectively. Let $v^i_0= \min\{u^i_0, \beta_i^- \}, i=1,...,N.$ Then $v_0=(v^i_0) \in \mathcal{C_{\beta^-}}$.  Letting
$u^{-}_n$ denote the $n$-th iteration of $\mathcal{Q}[F^-]$ starting from $v_0$ and observing that $v_0 \leq u_0$ and $\beta^- \leq \beta \leq \beta^+$, from (H2), we have that
$$
u_n^{-}(x) \leq u_n(x) \leq u_n^{+}(x), x \in \mathbb{R}, n >0.
$$

Theorem \ref{th20} (ii) states that for any strictly positive constant $\omega$, there is a positive $R_{\omega}$
(choose the larger one between the $R_{\omega}$ for $F^{+}$ and the $R_{\omega}$ for $F^{-}$) with the property that if $u_0 \geq \omega $
on an interval of length $2R_{\omega}$. Hence, it follows that  the solutions $u^{\pm}_n(x)$ satisfy
$$
\liminf_{t \to \infty} \inf_{\abs{x}\leq tc} u^{\pm}(x) =  \beta^{\pm}, \text{ for } 0< c< c^*.
$$
Thus for any $c<c^*$, it follows from Theorem \ref{th20} (ii) that $$
\liminf_{n \to \infty} \inf_{\abs{x}\leq nc} u^{\pm}_n(x) = \beta^{\pm},
$$
and consequently, that
$$
\beta^- \leq \liminf_{n \to \infty} \inf_{\abs{x}\geq nc} u_n(x) \leq  \beta^+.
$$
\epf

\section{Characterization of $c^*$ as the slowest speeds of traveling waves}\label{proofMono}
A non-constant solution of (\ref{eq1}) is a traveling wave of speed $c$ provided that it has the form $u_n(x)=u(x-cn)$,
where $u\in C(\mathbb{R}, \mathbb{R}^N)$ and, of course, if it satisfies  (\ref{eq1}).
By substituting this form into (\ref{eq1}), it follows that $u(\xi)$ must satisfy the following system of equations.
\begin{equation}\label{eq211}
\begin{split}
u(\xi)&=\mathcal{Q}_c[F(u)](\xi)=(\mathcal{Q}_c^i[F(u)](\xi)):=\mathcal{Q}[F(u)](\xi+c)\\
 \end{split}
\end{equation}
In this section we complete the proof of Theorem \ref{th30} (iii), (iv) and (v), that is, the portion of our main result
 that characterizes the spread speed $c^*$ as {\it the speed} of the slowest member of a family of non-constant traveling wave solutions.
 This is an extension of prior results for  monotone operators \cite{Weinberger2005} and for
 scalar equations \cite{Weinberger1978,Weinberger1982,hsu2008,LiLewis2009}.

\subsection{Upper and lower solutions}\label{upperlower}
In this subsection, we shall verify that $\phi^+$ and $\phi^-$ defined below are the upper and lower solutions of (\ref{eq211}) respectively.
These solutions are only continuous on $\mathbb{R}$. Upper and lower solutions of this type have been frequently used
 in the literature (see Diekmann \cite{Diekmann1978JMB}, Weinberger \cite{Weinberger1978}, Lui \cite{Lui1989}, Weinberger, Lewis and Li \cite{Weinberger2002-1},
  Rass and  Radcliffe \cite{Rass2003}, Weng and Zhao \cite{Weng2006}, more recently by Ma \cite{ma2007} and Wang \cite{Hwang2009}).
  In particular, the explicit use of upper vector-valued solutions can be traced to the work in \cite{Lui1989,Weinberger2002-1,Rass2003,Weng2006};
   for lower vector-valued solutions, in the context of multi-type epidemic models, to the work in \cite{Rass2003}; and in \cite{Weng2006}
   in the context of multi-type SIS epidemic models. Our construction of $\phi^+$ and $\phi^-$, the upper and lower solutions of (\ref{eq211}),
    is motivated by the research in these references.

 Our verification of the lower and upper solutions for $n$-dimensional systems is new and different from the above mentioned references. The details follow below.


Let $c>c*$,  $1< \gamma <2$, $q>1$  and recall the definitions of $\Lambda_{c}$ and  $\gamma \Lambda_{c}$ as utilized in Proposition \ref{lmeigen-1}. The
corresponding positive eigenvectors $\nu_{\Lambda_c}$ and $\nu_{\gamma \Lambda_c}$ of $B_{\mu}$ for the
eigenvalues $\lambda_{\mu}$ when $\mu=\Lambda_{c}, \gamma \Lambda_{c}$ can therefore be identified.

Define
$$
\phi^+(\xi)=(\phi^+_i),
$$
where
$$
\phi^+_i=\min\{\beta_i, \nu^i_{\Lambda_c}e^{-\Lambda_{c}\xi}\},\; \xi \in \mathbb{R};
$$
and
$$
\phi^-(\xi)=(\phi^-_i),
$$
where
$$
\phi^-_i=\max\{0, \nu^i_{\Lambda_c}e^{-\Lambda_c \xi }-q \nu^i_{\gamma \Lambda_c}e^{-\gamma \Lambda_c \xi}\},\; \xi \in \mathbb{R}.
$$
It is clear that if $\xi \leq  \frac{\ln \frac{\beta_i}{\nu^i_{\Lambda_c}}}{-\Lambda_{c}}$ then $\phi^+_i(\xi)=\beta_i$; and  if $\xi > \frac{\ln \frac{k_i}{\nu^i_{\Lambda_c}}}{-\Lambda_{c}}$ then $\phi^+_i(\xi)=\nu^i_{\Lambda_c}e^{-\Lambda_{c}\xi}$. Similarly, if  $\xi \leq \ln( q\frac{\nu^i_{\gamma \Lambda_c}}{\nu^i_{\Lambda_c}})\frac{1}{(\gamma -1) \Lambda_{c}}$ then $\phi^-_i(\xi)=0$; and if
$\xi > \ln( q\frac{\nu^i_{\gamma \Lambda_c}}{\nu^i_{\Lambda_c}})\frac{1}{(\gamma -1) \Lambda_{c}}$ then $\phi^-_i(\xi)=\nu^i_{\Lambda_c}e^{-\Lambda_c \xi }-q \nu^i_{\gamma \Lambda_c}e^{-\gamma \Lambda_c \xi}.$

We choose $q>1$ large enough so that
$$
\frac{\ln( q\frac{\nu^i_{\gamma \Lambda_c}}{\nu^i_{\Lambda_c}})}{(\gamma -1) \Lambda_{c}}>\frac{\ln \frac{\beta_i}{\nu^i_{\Lambda_c}}}{-\Lambda_{c}}
$$
and therefore
$$
\phi^+_i(\xi) > \phi^-_i(\xi), \xi \in \mathbb{R}.
$$

We verify in the two lemmas below that $\phi^+$ and $\phi^-$ are upper and lower solutions of (\ref{eq211}) respectively. It is assumed that Lemma \ref{upper} is valid when $F$ is monotone. In this case, $F^{\pm}=F, \beta^{\pm}=\beta$.

\begin{lemma}\label{upper} Assume $F$ is monotone and $(H1)-(H3)$ hold. For any $c > c^*$, then $\phi^+$ is an upper solution of $\mathcal{Q}_c[F]$. That is
$$
\mathcal{Q}_c[F(\phi^+)](\xi)\leq \phi^+(\xi), \xi \in \mathbb{R}.
$$
\end{lemma}
\pf
Let $\xi^*_i= \frac{\ln \frac{\beta_i}{\nu^i_{\Lambda_c}}}{\Lambda_{c}}.$ Then $\phi^+_i(\xi)=\beta_i$ if $\xi \leq \xi^*_i$, and $\phi^+_i(\xi)=\nu^i_{\Lambda_c}e^{-\Lambda_{c}\xi}$ if $\xi > \xi^*_i.$
Note that $\phi^+_i(\xi) \leq \nu^i_{\Lambda_c}e^{-\Lambda_{c}\xi}, \xi \in \mathbb{R}.$

In view of (H3) we have, for $\xi  \in \mathbb{R}$
\begin{equation*}
\begin{split}
f_i(\phi^+(\xi)) & \leq \sum_{j=1}^N\partial_j f_i(0) \phi_i^+(\xi) \leq \sum_{j=1}^N\partial_j f_i(0)\nu^j_{\Lambda_c}e^{-\Lambda_{c}\xi}
\end{split}
\end{equation*}
Thus, for $\xi \in \mathbb{R}$, in view of (\ref{matrix1}), (H3), Proposition \ref{lmeigen-1}, we obtain that
\begin{equation}\label{eq126}
\begin{split}
\mathcal{Q}^i[F(\phi^+)](\xi+c)& \leq e^{-\Lambda_c(\xi+c)}\sum_{j=1}^N\nu^j_{\Lambda_c} b_{\Lambda_c}^{i,j}\\
                            &=  e^{-\Lambda_c(\xi+c)} \lambda(\Lambda_c)\nu^i_{\Lambda_c}\\
                            & =  e^{-\Lambda_c(\xi+c)} e^{\Lambda_c \Phi(\Lambda_c)}\nu^i_{\Lambda_c}\\
                            & = \nu^i_{\Lambda_c} e^{-\Lambda_c (\xi+c)} e^{\Lambda_c c}\\
                            & = \nu^i_{\Lambda_c} e^{-\Lambda_c \xi}.
\end{split}
\end{equation}
On the other hand, since $\phi^+_i(\xi) \leq \beta_i, i=1,...,N$ , we have for $\xi \in \mathbb{R}$
\begin{equation}\label{eq128}
\begin{split}
\mathcal{Q}^i[F(\phi^+)](\xi+c)& \leq \beta_i.
\end{split}
\end{equation}
Thus,
we have for $\xi \in \mathbb{R}$
\begin{equation}\label{eq145}
\begin{split}
\mathcal{Q}_c^i[F(\phi^+)](\xi)=\mathcal{Q}^i[F(\phi^+)](\xi+c)& \leq \phi^+_i(\xi).
\end{split}
\end{equation}

This completes the proof of Lemma \ref{upper}.

\epf

In order to verify  the lower solution,  the following estimate for $F$ is needed. For $N=1,2$, Lemma \ref{estimageg} can be found in \cite{Hwang2009}.
\begin{lemma}\label{estimageg} Assume $(H1-H2)$ hold. There exist positive constants $D_{i}, i=1,...,N$ such that
$$
f_i(u)\geq \sum_{j=1}^N\partial_j f_i(0)u^j- D_{i}\sum_{j=1}^N(u^j)^2, \;\; u=(u^j), u \in [0,\beta^{+}], i=1,...,N.
$$
\end{lemma}
\pf
In a sufficiently small neighborhood of the origin, since $F$ is twice continuously differentiable. From the Taylor's Theorem for multi-variable functions
(the big Oh notation version), for $u$ sufficiently small.
$$
f_i(u)= \sum_{j=1}^N\partial_j f_i(0)u^j+ O(\sum_{j=1}^N(u^j)^2), \;\; u=(u^j), u \in [0,\beta^{+}], i=1,...,N.
$$
There exist small $\epsilon>0$ and $D'_{i}>0$ such that for $\sum_{j=1}^n (u^j)^2 < \epsilon$
$$
f_i(u)\geq \sum_{j=1}^N\partial_j f_i(0)u^j- D'_i \sum_{j=1}^N(u^j)^2, \;\; u=(u^j), u \in [0,\beta^{+}], i=1,...,N.
$$
For $u \in [0, \beta]$ and $\sum_{j=1}^n (u^j)^2 \geq  \epsilon$,  noting that $f_i(u), \sum_{j=1}^N\partial_j f_i(0)u^j$ are bounded,
we always choose a sufficiently large constant $D''_i>0$ such that
$$f_i(u)\geq \sum_{j=1}^N\partial_j f_i(0)u^j- D''_i\sum_{j=1}^N(u^j)^2.$$
Thus if we let $D_{i}=\max\{D'_i, D''_i\}$,  then Lemma \ref{estimageg}
is proved.
\epf

\begin{lemma}\label{sub} Assume $(H1)-(H3)$ hold. For any $c > c^*$  if $q$ (which is independent of $\xi$) and that it is sufficiently large,
$\phi^-$ is a lower solution of $\mathcal{Q}_c[F]$. That is
$$
\mathcal{Q}_c[F(\phi^-)](\xi) \geq \phi^-(\xi), \;\;\xi \in \mathbb{R}.
$$
\end{lemma}
\pf Again let $\xi^*_i=\ln( q\frac{\nu^i_{\gamma \Lambda_c}}{\nu^i_{\Lambda_c}})\frac{1}{(\gamma -1) \Lambda_{c}}$. Hence if  $\xi \leq \xi^*_i$ then
$\phi_i^-(\xi)=0$; while  if $\xi > \xi^*_i$ then $\phi_i^-(\xi)=\nu^i_{\Lambda_c}e^{-\Lambda_c \xi }-q \nu^i_{\gamma \Lambda_c}e^{-\gamma \Lambda_c \xi}$.
It is easy to see that
\begin{equation}\label{eq250}
\begin{split}
\nu^i_{\Lambda_c}e^{-\Lambda_{c}\xi} \geq  \phi^-(\xi) \geq \nu^i_{\Lambda_c}e^{-\Lambda_c \xi }-q \nu^i_{\gamma \Lambda_c}e^{-\gamma \Lambda_c \xi}, \;\; \xi \in \mathbb{R}, i=1,...,N.
\end{split}
\end{equation}
For $\xi \in \mathbb{R}$, in view of  Lemma \ref{estimageg}, we have, for $\xi \in \mathbb{R}, i=1,...,N$,
\begin{equation}\label{eq31}
\begin{split}
&f^i(\phi^-(\xi)) \geq \sum_{j=1}^N\partial_j f_i(0)\phi_j^-(\xi) -D_{i}\sum_{j=1}^N (\phi_j^-(\xi))^2\\
             & \geq \sum_{j=1}^N\partial_j f_i(0)\nu^j_{\Lambda_c} e^{-\Lambda_c \xi} - q \sum_{j=1}^N\partial_j f_i(0)\nu^j_{\gamma \Lambda_c} e^{-\gamma \Lambda_c \xi} -\widehat{M}_i e^{-2\Lambda_{c}\xi}\\
\end{split}
\end{equation}
where $
\widehat{M}_i=D_{i}\sum_{j=1}^N(\nu^j_{\Lambda_c})^2>0.
$
Now we are able to estimate $\mathcal{Q}[\phi^-]$
for  $\xi \geq \min_i\xi^*_i, i=1,...,N$ as in (\ref{eq126})
\begin{equation}\label{eq347}
\begin{split}
\mathcal{Q}^i[F(\phi^-)](\xi+c)& \geq e^{-\Lambda_c(\xi+c)}\sum_{j=1}^N\nu^j_{\Lambda_c} b_{\Lambda_c}^{i,j} -q e^{-\gamma \Lambda_c(\xi+c)}\sum_{j=1}^N\nu^j_{\gamma \Lambda_c} b_{\gamma\Lambda_c}^{i,j}\\
                      &\quad-\widehat{M}_i e^{-2 \Lambda_c(\xi+c)} \int_{\mathbb{R}}k_i(y)e^{2\Lambda_cy }dy\\
                      &= \nu^i_{\Lambda_c}e^{-\Lambda_c(\xi+c)}e^{\Lambda_c \Phi(\Lambda_c)}-q \nu^i_{\gamma \Lambda_c}e^{-\gamma \Lambda_c(\xi+c)}e^{\gamma \Lambda_c \Phi(\gamma \Lambda_c)}\\
                      &\quad-\widehat{M}_i e^{-2 \Lambda_c(\xi+c)} \int_{\mathbb{R}}k_i(y)e^{2\Lambda_cy}dy\\
                      &= \nu^i_{\Lambda_c}e^{-\Lambda_c \xi }-q \nu^i_{\gamma \Lambda_c}e^{-\gamma \Lambda_c \xi} e^{\gamma \Lambda_c (\Phi(\gamma \Lambda_c)-c)}\\
                      &\quad-\widehat{M}_i e^{-2 \Lambda_c(\xi+c)} \int_{\mathbb{R}}k_i(y)e^{2\Lambda_cy}dy\\
                      &= \nu^i_{\Lambda_c}e^{-\Lambda_c \xi }-q \nu^i_{\gamma \Lambda_c}e^{-\gamma \Lambda_c \xi}\\
                      &\quad +q \nu^i_{\gamma \Lambda_c}e^{-\gamma \Lambda_c \xi} -q \nu^i_{\gamma \Lambda_c}e^{-\gamma \Lambda_c \xi} e^{\gamma \Lambda_c (\Phi(\gamma \Lambda_c)-c)}\\
                      &\quad-\widehat{M}_i e^{-2 \Lambda_c(\xi+c)} \int_{\mathbb{R}}k_i(y)e^{2\Lambda_cy}dy\\
                      & =\phi^-_i(\xi) + e^{-\gamma \Lambda_c \xi}\Big(q \nu^i_{\gamma \Lambda_c}\big(1 -e^{\gamma \Lambda_c (\Phi(\gamma \Lambda_c)-c)} \big)\\
                      &\quad -\widehat{M}_i e^{(\gamma -2)\Lambda_c \xi}e^{-2 \Lambda_c c} \int_{\mathbb{R}}k_i(y)e^{2\Lambda_cy}dy\Big)
\end{split}
\end{equation}
For $\xi \geq \min_i\xi^*_i,$ $e^{(\gamma-2)\Lambda_{c}\xi}$ is bounded above.
Finally, from (\ref{eq347}) and the fact that $\Phi(\gamma \Lambda_c)<c$, we conclude that there exists $q>0$, which is independent of $\xi$, such that, for $\xi \geq \xi^*_i$
\begin{equation}\label{eq37}
\begin{split}
\mathcal{Q}^i[F(\phi^-)](\xi+c)                      & \geq \nu^i_{\Lambda_c}e^{-\Lambda_c \xi }-q \nu^i_{\gamma \Lambda_c}e^{-\gamma \Lambda_c \xi}.
\end{split}
\end{equation}
And since $\phi_i^-(\xi)=0$ for $\xi<\xi^*_i$, $i=1,...,N$
\begin{equation*}
\begin{split}
\mathcal{Q}_c^{i}[F(\phi^-)](\xi)=\mathcal{Q}^i[F(\phi^-)](\xi+c)                      & \geq \phi_i^-(\xi), \;\; \xi \in \mathbb{R}.
\end{split}
\end{equation*}
This completes the proof.
\epf

\subsection{Proof of Theorem \ref{th30} (iii) with monotonicity of $F$}\label{monotoproof}
Theorems that guarantee the existence of traveling wave solutions for cooperative systems have been established (e.g. \cite{Weinberger2005,Weinberger1982}).
In this section, it is assumed that $F$ is non-decreasing on $[0,\beta]$ and from this assumption, we proceed to establish Theorem \ref{th30}.

As we state in Section \ref{rest}, even for the case of monotone operators,  the results and analysis in this manuscript are different from \cite{Weinberger2005,Weinberger1982}. Here, we are able to characterize explicitly the asymptotic behavior of traveling waves through a careful analysis of eigenvalues and upper-lower solutions ( an analysis not provided in \cite{Weinberger2005,Weinberger1982}). As we shall see, the analysis of the asymptotic behavior of traveling wave solutions for monotone operator enable us also   {\it  to prove the existence of traveling wave solutions for non monotone operators}.

In order to complete the last step, we need to make use of the following Banach space
$$
\mathcal{B}_{\rho}= \{u=(u^i): u^i \in C(\mathbb{R}),\;\;\; \sup_{ \xi\in \mathbb{R}} \abs{u^i(\xi)}e^{\rho \xi} < \infty, i=1,...,N\},
$$
equipped with the weighted norm
$$
\norm{u}_{\rho}=\sum_{i=1}^N\sup_{ \xi\in \mathbb{R}} \abs{u^i(\xi)}e^{\rho \xi},
$$
where $C(\mathbb{R})$ denotes the set of all continuous functions on $\mathbb{R}$,  and where  $\rho$ is a positive constant such that $\rho<\Lambda_{c}.$
It follows that $\phi^+ \in \mathcal{B}_{\rho}$ and $\phi^-\in \mathcal{B}_{\rho}.$
Finally, the following set is required (domain of the operator of interest):
$$
\mathcal{A}_{\rho}=\{ u: u \in \mathcal{B}_{\rho}, \phi^-(\xi) \leq u(\xi) \leq \phi^+(\xi), \xi \in \mathbb{R}\}
$$
It is clear that $\mathcal{A}_{\rho} \subseteq \mathcal{C}_{\beta}$. By the standard procedure (see \cite{ma2007,hsu2008,Hwang2009}),
it can be shown that
$\mathcal{Q}_c[F]$ is a continuous map of the bounded set
$\mathcal{A}_{\rho}$ into a compact set.

\begin{lemma}\mlabel{conunity} Assume $(H1)-(H3)$ hold. Then
$\mathcal{Q}_c[F]: \mathcal{A}_{\rho} \rightarrow \mathcal{A}_{\rho}$ is continuous with the weighted norm $\norm{.}_{\rho}$ and relatively compact in $\mathcal{B}_{\rho}.$
\end{lemma}

Now we are in a position to prove Theorem \ref{th30} when $F$ is monotone. Define the following iteration
\begin{equation}\label{iteration1}
u_1=(u_1^i)=\mathcal{Q}_c[F(\phi^+)], \;\; u_{n+1}=(u^i_n)=\mathcal{Q}_c[F(u_n)], n \geq 1.
\end{equation}
From Lemmas \ref{upper}, \ref{sub}, and the fact that $F$ is non-decreasing, $u_n$ is non-increasing on $\mathbb{R}$, it follows that $$
\phi_i^-(\xi) \leq u^i_{n+1}(\xi) \leq u^i_n(\xi) \leq \phi_i^+(\xi), \xi \in \mathbb{R}, \;n \geq 1,i=1,...,N.
$$
By Lemma \ref{conunity} and monotonicity of ($u_n$), there is $u \in \mathcal{A}_{\rho}$ such that $\lim_{n\to \infty}\norm{u_n-u}_{\rho}=0$.  Lemma \ref{conunity}
implies that $\mathcal{Q}[u]=u$. Furthermore, $u$ is non-increasing. It is clear that $\lim_{\xi \to \infty}u^i(\xi)=0,i=1,...,N$.
Assume that $\lim_{\xi \to -\infty}u_i(\xi)=\hat{k^i},i=1,...,N$ $\hat{k^i}>0, i=1,...,N$ because of $u \in \mathcal{A}_{\rho}$.
Applying the dominated convergence theorem, we get $\hat{k_i}=f_i(\hat{k}).$ By (H2), $\hat{k}=\beta$. Finally, note that
$$
\nu^i_{\Lambda_c}(e^{-\Lambda_{c} \xi}-q\e^{-\gamma \Lambda_{c}\xi}) \leq u^i(\xi) \leq \nu^i_{\Lambda_c}e^{-\Lambda_{c}\xi}, \xi \in \mathbb{R}.
$$
We immediately obtain that
\begin{equation}\label{aystom1}
\lim_{\xi \to \infty}u^i(\xi)e^{\Lambda_{c}\xi}=\nu^i_{\Lambda_c}, i=1,...,N.
\end{equation}
This completes the proof of Theorem \ref{th30} when $F$ is monotone.

\subsection{Proof of Theorem \ref{th30} (iii)} \label{proofofTh1iii}
We proceed to characterize traveling wave solutions when the assumption that $F$ is monotone is {\it dropped}. Our treatment is
different even for the scalar case ($N=1$). The key mathematical ideas used can be found in the literature albeit there are differences. Our use of the Schauder Fixed Point Theorem and the construction of the bounded set  $\mathcal{D}_{\rho}$ are different from those found in \cite{hsu2008,LiLewis2009}.
As in Section \ref{proofMono}, both $\mathcal{Q}_c[F^+]$ and $\mathcal{Q}_c[F^-]$ are monotone.  Note that $F, F^+, F^-$ have the same linearization at the origin. In view of the results in Section \ref{proofMono}, there exists a non-increasing
fixed point $u_-=(u_-^i) \in \mathcal{C}_{\beta^-}$ of $\mathcal{Q}_c[F^-]$ such that
$$\mathcal{Q}_c[F^-(u_-)]=u_-$$
and
$\lim_{\xi \to -\infty}u^i_{-}(\xi)=\beta^-_i, i=1,...,N$, and $\lim_{\xi \to \infty}u^i_{-}(\xi)=0, i=1,...,N$. Furthermore,
$\lim_{\xi \to \infty}u^i_{-}(\xi)e^{\Lambda_{c}\xi}=\nu^i_{\Lambda_c}, i=1,...,N.$
Let
$$
\widetilde{\phi^+}(\xi)=(\widetilde{\phi^+_i}),
$$
where
$$
\widetilde{\phi^+_i}=\min\{\beta_i^+, \nu^i_{\Lambda_c}e^{-\Lambda_{c}\xi}\},\; \xi \in \mathbb{R},i=1,...,N.
$$
According to Lemma \ref{upper}, $\widetilde{\phi^+}$ is an upper solution of $\mathcal{Q}_c[F^+]$. Also note that if $\beta^+$ is replaced with $\beta^-$,
$\widetilde{\phi^+}(\xi)$ is an upper solution of $\mathcal{Q}_c[F^-]$.  By the construction of $u^i_-(\xi)$, it then follows that
$$u_-(\xi) \leq \widetilde{\phi^+}(\xi), \xi \in \mathbb{R}$$
Now let
$$
\mathcal{D}_{\rho}=\{ u: u=(u^i) \in \mathcal{B}_{\rho}, u^i_-(\xi) \leq u^i(\xi) \leq \widetilde{\phi^+_i}(\xi), \xi \in ( -\infty, \infty), i=1,...,N\},
$$
where $\mathcal{B}_{\rho}$ is defined in Section \ref{monotoproof}.
It is clear that $\mathcal{D}_{\rho}$ is a bounded nonempty closed convex subset in $\mathcal{B}_{\rho}$.
Furthermore, we have, for any $u=(u^i) \in \mathcal{D}$
$$
u_- = \mathcal{Q}_c[F^-(u_-)]\leq \mathcal{Q}_c[F^-(u)] \leq \mathcal{Q}_c[F(u)] \leq \mathcal{Q}_c[F^+(u)]\leq \mathcal{Q}_c[F^+(\widetilde{\phi^+})] \leq \widetilde{\phi^+}.
$$
Therefore,  $\mathcal{Q}_c[F]: \mathcal{D}_{\rho} \rightarrow \mathcal{D}_{\rho}$. Note that the proof of Lemmas \ref{conunity} does not
need the monotonicity of $F^{-}$. In the same way as in Lemmas \ref{conunity}, we can show that
$\mathcal{Q}_c[F^-]: \mathcal{D}_{\rho} \rightarrow \mathcal{B}_{\rho}$ is continuous and maps bounded sets into compact sets.
Therefore, the Schauder Fixed Point Theorem shows that the operator $\mathcal{Q}_c[F]$ has a fixed point $u$ in $\mathcal{D}_{\rho}$,
which is a traveling wave solution of (\ref{eq1}) for $c>c^*$. Since $u_i^-(\xi) \leq u^i(\xi) \leq \widetilde{\phi^+_i}(\xi), \xi \in ( -\infty, \infty), i=1,...,N$,
it is easy to see that for $i=1,...,N$,
$\lim_{\xi \to \infty}u^i(\xi)=0$, $\lim_{\xi \to \infty}u^i(\xi)e^{\Lambda_{c}\xi}=\nu^i_{\Lambda_c}$,
$$\beta^-_i\leq \liminf_{\xi \to -\infty}u^i(\xi) \leq \limsup_{\xi \to -\infty}u^i(\xi)\leq \beta_i^+$$
and
$0 < u^i_-(\xi) \leq  u^i(\xi) \leq \beta_i^+, \xi \in ( -\infty, \infty)$.
\epf

\subsection{Proof of Theorem \ref{th30} (iv)} \label{proofofTh1iv}
\pf The proof in this subsection follows the approach found in \cite{BrownCarr1977,hsu2008}. We make use of the results in Theorem \ref{th30} (iii). Hence, for each $m \in \mathbb{N}$, we choose $c_m >c^*$ such that $\lim_{m\to \infty} c_m=c^*.$ According to Theorem \ref{th30} (iii), for each $c_m$ there is a traveling wave solution $u_m=(u^i_m)$ of (\ref{eq1}) such that
$$
u_m=\mathcal{Q}[F(u_m)](\xi+c_m).
$$
and
$$\lim_{\xi \to \infty}u^i(\xi)=0,\; \beta^-_i\leq \liminf_{\xi \to -\infty}u^i_m(\xi) \leq \limsup_{\xi \to -\infty}u^i_m(\xi)\leq \beta_i^+, i=1,...,N.$$
By the standard procedure (see \cite{ma2007,hsu2008,Hwang2009}), $(u_m)$ is equicontinuous and uniformly bounded on $\mathbb{R}$. Hence, the Ascoli's theorem implies that there is a vector valued continuous function $u=(u^i)$ on $\mathbb{R}$ and subsequence $(u_{m_k})$ of $(u_m)$,  such that $$
\lim_{k \to \infty} u_{m_k}(\xi)=u(\xi)
$$
uniformly in $\xi$ on any compact interval of $\mathbb{R}$.  Further, the use of the dominated convergence theorem guarantees that we have
$$
u=\mathcal{Q}[F(u)](\xi+c^*)
$$
Because of the translation invariance of $u_m$, we always can assume that the first component $u_m^1(0)$ equals to a sufficiently small positive number $\sigma>0$ for all $m$.  Since there is only a finite number of equilibria,  we can choose $\sigma$ in such a way that it is not the first component of any nontrivial equilibrium.  Consequently $u$ is a nonconstant traveling solution of (\ref{eq1}) for $c=c^*$.
\epf

\subsection{Proof of Theorem \ref{th30} (v)} \label{proofofTh1v}
The proof of this subsection follows the approach in \cite{hsu2008,LiLewis2009}. Suppose, by contradiction, that for some $c \in (0, c^*)$, (\ref{eq1}) has a traveling wave $u_n(x)=u(x - cn)$
such that $u \in \mathcal{C}_{\beta} $ with $ \liminf_{x \to -\infty} u(x) \gg 0$ and $u(+\infty)=0.$ Thus $u(x)$ can be larger than
a positive vector with arbitrary length. It follows
from Theorem \ref{th30} (ii)
$$
\liminf_{n \to \infty} \inf_{\abs{x}  \leq nc} u_n(x) \geq \beta^-, \text{ for } 0< c< c^*
$$
Let $\hat{c} \in (c,c^*)$ and $x=\hat{c}n.$ Then $$
\lim_{n\to \infty} u\big((\hat{c}-c)n\big) =\lim_{n\to \infty} u_n(\hat{c}n) \geq \liminf_{n \to \infty} \inf_{\abs{x}  \leq n\hat{c}} u_n(x) \geq \beta^-.
$$
However,
$
\lim_{n \to \infty}u\big((\hat{c}-c)n\big)=u(\infty)=0,
$
which is a contradiction.
\epf

\section{Minimum speeds and traveling waves for a competition model}\label{example}
Hassell and Comins' model of the growth and spread of two population densities at time $n$ and location $x$ under an interference competition regime is used to highlight the applicability of the results in this manuscript. We makes use of the local analysis results of their model reported in  \cite{Hassell}. The addition of the possibility of  dispersal via the re-distribution kernel $k_i(x-y)$ leads to (\ref{eqoper01}).   If their two densities are  denoted by  $X_n(x)$ and $Y_n(x)$ then their model is given locally by a set of nonlinear coupled difference equations (\ref{eqoper011}) with the addition of dispersal leading to  (\ref{eqoper01}).  The following results highlight the contributions that the main theorem makes towards increasing our understanding of the role of dispersal, in the context of local competitive systems. Li \cite{Li2010} also investigated the minimum speed of (\ref{eqoper01}).

Model (\ref{eqoper01}) can support four constant equilibria: the unpopulated state $(0,0);$ the second-species monoculture state $(0,r_2);$
the first monculture state $(r_1,0);$ and $(\frac{r_1-\sigma_1r_2}{1-\sigma_1\sigma_2}, \frac{r_2-\sigma_2r_1}{1-\sigma_1\sigma_2}).$ The change of variables $
p=X, q = r_2-Y
$
allows to convert system (\ref{eqoper01}) into the following coupled system of integro-difference equations
\begin{equation}\label{eqoper2}
\begin{split}
p_{n+1}(x)&=\int_{\mathbb{R}}k_1(x-y)f(p_n(y),q_n(y))dy\\
q_{n+1}(x)&=\int_{\mathbb{R}}k_2(x-y)g(p_n(y),q_n(y))dy.
\end{split}
\end{equation}
where
\begin{equation*}
\begin{split}
f(p,q)&=h(p)e^{r_1-\sigma_1 r_2+\sigma_1 q}\\
g(p,q)&=r_2-\big(r_2-q\big)e^{q-\sigma_2 p}\\
h(p)&=pe^{-p}
\end{split}
\end{equation*}
It is clear that (\ref{eqoper01}) and (\ref{eqoper2}) are not monotone systems. A straightforward calculation shows that (\ref{eqoper2}) has
four equilibria $(0,0), (0,r_2), (r_1,r_2)$ and

\begin{equation*}
(\frac{r_1-\sigma_1r_2}{1-\sigma_1\sigma_2}, \sigma_2\frac{r_1-\sigma_1r_2}{1-\sigma_1\sigma_2}).
\end{equation*}
In fact, under the conditions of Theorem \ref{th33},  we show in Appendix that there are no positive equilibrium of (\ref{eqoper2})
between $(0,0)$ and $(r_1, r_2)$. Theorem \ref{th30} is  used to guarantee the existence of a spreading speed and traveling wave solutions
of the nonmonotone system (\ref{eqoper2}) with its accompanying results on the speed of propagation..
We summarize the results obtained in the context of this example in Theorem \ref{th33}. Its proof is outlined  in the Appendix.

\begin{theorem}\label{th33}  Let $0<r_2<1<r_1$, $0<\sigma_1<1< \sigma_2, \sigma_1\sigma_2<1$, and $$r_2< \sigma_2 e^{r_1-1-e^{r_1-1}}$$
and $$\sigma_1 r_2< e^{r_1-1-e^{r_1-1}}.$$ Assume that $k_1, k_2$ satisfy
(H1) and
$\int_{\mathbb{R}}k_1(s)e^{\mu s}ds \geq \int_{\mathbb{R}}k_2(s)e^{\mu s}ds$ for $\mu >0$.
Then the conclusions of Theorem \ref{th30} hold for (\ref{eqoper2}).
\end{theorem}
The biological interpretation of the conditions in Theorem \ref{th33} in the context of our application are straightforward.
For an invasion to be successful, the overall dispersal of the invader (X) is relatively larger than the overall dispersal of the out-competed resident (Y).  Further
competition favors the invader whenever $\sigma_1$ is sufficiently small (invader less affected by competition) and $\sigma_2$ is sufficiently large (a relatively fragile resident, that is, more susceptible to interference competition). Under these conditions, there are traveling wave solutions of (\ref{eqoper2}) ``loosely" connecting its two equilibria $(0,0)$ and $(r_1,r_2)$. Equivalently, there are traveling wave solutions of (\ref{eqoper01}) ``loosely" connecting its two boundary states $(0,r_2)$ and $(r_1,0)$. Here the term ``loosely" means the traveling waves may oscillate around the equilibria since they are
not necessarily monotone. For specific $k_i$, the exact value of $c^*$ can be computed and compared to experimental data as it has been done by Kot, Lewis, others and their collaborators.

\section{Conclusions}

integro-difference systems arise naturally in the study of the dispersal of populations, including interacting populations, composed of organisms that reproduce locally via discrete generations and compete for resources, before dispersing .  The brunt of the {\it mathematical} research has focused on the the study of the existence of traveling wave solutions and characterizations of the spreading speed in the context of cooperative systems. In this paper, we characterize the spreading speed for a large class of {\it non cooperative systems}, formulated in terms of integro-difference equations, via the convergence of initial data to wave solutions. The spreading speed is characterized as the slowest speed of a family of non-constant traveling wave solutions. The results are applied to  the {\it non-cooperative} competitive system proposed by Hassell and Comins (1976) \cite{Hassell}. We are in the process of applying these results to additional ecological and epidemiological systems where the local dynamics are naturally non-cooperative with the hope that increasing our understanding of the role of dispersal in communities where the local dynamics are richer, more realistic, than those previously supported by the mathematical theory.

\section*{Acknowledgements}
This project has been partially supported by grants from the National Science Foundation (NSF - Grant DMPS-0838704), the National Security Agency (NSA - Grant H98230-09-1-0104), the Alfred P. Sloan Foundation and the Office of the Provost of Arizona State University.
The authors would thank the reviewers for their carefully reading of the manuscript and constructive comments.
\section*{Appendix}
\textbf{Proof} of Proposition \ref{lmeigen-1} (4).

If $\Phi(\mu)=\frac{1}{\mu}\ln \lambda(\mu)$ achieves its minimum at a finite $\mu$, then $c^*=\min_{\mu >0} \Phi(\mu)>0$.
Now let $c^*= \lim_{\mu \to \infty} \frac{1}{\mu} \ln \lambda(\mu)$. We recall that $\lambda(\mu)$ is
an eigenvalue of $B_{\mu}$ with a positive eigenvector. Thus there
exists a positive constant $\delta>0$ and a positive integer $i \leq N $ such that
$
\lambda(\mu) \geq \delta \int_{\mathbb{R}} k_i(x)e^{\mu x}dx.
$
Thus
$$
c^* \geq \lim_{\mu \to \infty} \frac{1}{\mu}\ln \big(\delta \int_{\mathbb{R}} k_i(x)e^{\mu x}dx\big).
$$
Let $
\Psi(\mu)=\frac{\int_{\mathbb{R}} x k_i(x)e^{\mu x}dx}{\int_{\mathbb{R}} k_i(x)e^{\mu x}dx},\;\; \mu \geq 0.
$
Then by the L'Hopital's rule  we have
$
c^* \geq \lim_{\mu \to \infty} \Psi(\mu).
$
Differentiation of $\Psi$ and rearrangement of terms show
$$
\Psi'(\mu)=\frac{\int_{\mathbb{R}}\big( x-\Psi(\mu)\big)^2 k_i(x)e^{\mu x}dx}{\int_{\mathbb{R}} k_i(x)e^{\mu x}dx}>0, \mu \geq 0,
$$
also see Weinberger \cite{Weinberger1978}. Note that $\Psi(0)=0$ and therefore, $c^* \geq \lim_{\mu \to \infty} \Psi(\mu)>0.$
\epf

\noindent \textbf{Proof} of Theorem \ref{th33}.

We verify that the conditions (H1-H3) hold for (\ref{eqoper2}).  From the assumptions of Theorem \ref{th33}, (H1) holds for (\ref{eqoper2}).  We proceed to verify (H2) for (\ref{eqoper2}) which, as we had noticed earlier, has
four equilibria $(0,0), (0,r_2), (r_1,r_2)$ and

\begin{equation}\label{eq9090}
(\frac{r_1-\sigma_1r_2}{1-\sigma_1\sigma_2}, \sigma_2\frac{r_1-\sigma_1r_2}{1-\sigma_1\sigma_2}).
\end{equation}
If it is further assumed that $r_1 >1, r_2<1$, and  $\sigma_1<1, \sigma_2>1, \sigma_1\sigma_2<1$ then
\begin{equation}\label{eq8080}
(\frac{r_1-\sigma_1r_2}{1-\sigma_1\sigma_2}, \sigma_2\frac{r_1-\sigma_1r_2}{1-\sigma_1\sigma_2})\gg (r_1,r_2).
\end{equation}
Thus (\ref{eqoper2}) has no other positive equilibrium $(\underline{p},\underline{q})$ between $(0,0)$ and $(r_1, r_2)$ with
$\underline{p}>0$ and $\underline{q}>0.$  Observe that $1$ is the maximum point of $h(p)$, that is, $h(p)$ is not monotone on $[0, r_1].$ Further simple calculations show that
$g_p(p,q)=\sigma_2(r_2-q)e^{q-\sigma_2p} \geq 0, \text{ for } q\in [0, r_2]$,
$g_q(p,q)=(1-r_2+q)e^{q-\sigma_2 p} \geq 0$.

In order to use Theorem \ref{th30}, we define the upper monotone function
\begin{equation*}
h^{+}(p) = \left\{ \begin{array}{ll}
h(p),    & \;\;\;\;\; 0 \leq p \leq 1, \\[.2cm]
h(1)=e^{-1}, & \;\;\;\;\; 1 \leq p.
\end{array} \right.
\end{equation*}
and corresponding monotone systems with $h^{+}$
\begin{equation}\label{eqoper2+}
\begin{split}
p_{n+1}(x)&=\int_{\mathbb{R}}k_1(x-y)f^+(p_n(y),q_n(y))dy\\
q_{n+1}(x)&=\int_{\mathbb{R}}k_2(x-y)g(p_n(y),q_n(y))dy.
\end{split}
\end{equation}
where
$
f^{+}(p,q)=h^{+}(p)e^{r_1-\sigma_1 r_2+\sigma_1 q}.
$

The origin, $(0,0)$ is an equilibrium  of  (\ref{eqoper2+}) and $g(p,q)=q$ has only two possible solutions $q^*=r_2$ and $q^*= \sigma_2p^*.$
Thus for $q^*=r_2$, Equation (\ref{eqoper2+}) has two equilibria  $(0,r_2), (e^{r_1-1},r_2)$. The second equilibrium $(e^{r_1-1},r_2)$ comes from
the fact that $p^* >1$ and therefore $h^{+}(p^*)=e^{-1}$.  (If $0<p^* \leq 1$, then $h^+(p^*)=h(p^*)$ and $(p^*,q^*)=(r_1,r_2)$;
however, $r_1>1$, which is a contradiction).  In order for Equation (\ref{eqoper2+}) to have another positive equilibrium $(p^*, q^*)$, when $q^*=\sigma_2p^*$,  it must satisfy $p^*>1$ (otherwise, $p^* \leq 1$ and $(p^*, q^*)$ is (\ref{eq9090}) which means that  (\ref{eq8080}) implies that $p^*>1$, a contradiction) and therefore
\begin{equation}\label{equilib256}
\begin{split}
e^{r_1-\sigma_1 r_2+\sigma_1\sigma_2 p^*-1} &=p^* \\
q^*&=\sigma_2 p^*.
\end{split}
\end{equation}
We will use the inequality, $e^x \geq x+1$, $x \in \mathbb{R}$  to estimate $e^x$. Thus $p^*= e^{r_1-\sigma_1 r_2+\sigma_1\sigma_2 p^*-1} \geq r_1-\sigma_1 r_2+\sigma_1\sigma_2 p^*$ and $p^* \geq \frac{r_1-\sigma_1 r_2}{1-\sigma_1\sigma_2} > r_1,$ which implies that $ p^*> e^{r_1-1}$, from the first equation of (\ref{equilib256}) and $\sigma_2r_1>r_2$. Again since $
\sigma_2r_1>r_2$, we also have $q^*=\sigma_2p^*>r_2$ and thus Equation (\ref{eqoper2+}) has no positive equilibrium between  $(0,0)$ and $(e^{r_1-1},r_2).$

There is a $t_0 \in (0, 1)$ such that $h(t_0)=h(e^{r_1-1})$ and define
\begin{equation*}
h^{-}(p) = \left\{ \begin{array}{ll}
h(p),    &\;\;\; 0 \leq p \leq t_0, \\[.2cm]
h(t_0), & \;\;\;t_0 \leq p \leq e^{r_1-1}.
\end{array} \right.
\end{equation*}
and corresponding lower monotone system
\begin{equation}\label{eqoper2-}
\begin{split}
p_{n+1}(x)&=\int_{\mathbb{R}}k_1(x-y)f^-(p_n(y),q_n(y))dy\\
q_{n+1}(x)&=\int_{\mathbb{R}}k_2(x-y)g(p_n(y),q_n(y))dy.
\end{split}
\end{equation}
where
$
f^{-}(p,q)=h^{-}(p)e^{r_1-\sigma_1 r_2+\sigma_1 q}.
$
Then $$0<h^-(p) \leq h(p) \leq  h^+(p) \leq h'(0)p, p \in (0, e^{r_1-1}]$$
$h^-(0)=h^+(0)=0,$ $h^{\pm}(p), h(p)$ have the same derivative at $0$.

Since $g(p,q)=q$ has only two possible solutions $q=r_2$ and $q=\sigma_2p$ and $h(t_0)=t_0e^{-t_0}=h(e^{r_1-1})=e^{r_1-1}e^{-e^{r_1-1}}$, we can therefore calculate that (\ref{eqoper2-}) has three equilibria
$(0,0), (0, r_2)$ and $(t_1,r_2)$ where $t_1=e^{2r_1-1 -e^{r_1-1}}$. Again $(t_1,r_2)$ comes from the fact that $t_1 \geq t_0$ and $h(t_1)=h(t_0)$.
(The same argument applied to (\ref{eqoper2+}) implies that $t_1 < t_0$ is a contradiction. ) We will now show that
\begin{equation}\label{eq555}
t_1<r_1.
\end{equation}
%
Indeed, since Expression (\ref{eq555}) is equivalent to $
2r_1-1 -e^{r_1-1}< \ln r_1,
$
we let  $l(x)=2x-1 -e^{x-1}- \ln x$ and therefore  $l(1)=0$ and
$$
l'(x)=2-e^{x-1}-\frac{1}{x} \leq 2-x-\frac{1}{x} <0, \text{ for } x>1,
$$
and this verifies (\ref{eq555}).
Since  $0<t_0<1<r_1$, the following inequality holds
\begin{equation}\label{eq9989}
\begin{split}
0<e^{r_1-1-e^{r_1-1}}<t_0=e^{t_0}e^{r_1-1}e^{-e^{r_1-1}} < e^{2r_1-1-e^{r_1-1}}= t_1.
\end{split}
\end{equation}
If $(p^*,q^*)$ is a another positive equilibrium of (\ref{eqoper2-}) when $q^*=\sigma_2p^*,$ then it must
satisfy $p^*>t_0$ (otherwise, $p^* \leq t_0$ and $(p^*, q^*)$ is (\ref{eq9090}) and
from (\ref{eq8080}) we have that $p^*>1>t_0$, a contradiction) and therefore
\begin{equation}\label{equilib259}
\begin{split}
e^{r_1-1}e^{-e^{r_1-1}}e^{r_1-\sigma_1 r_2+\sigma_1\sigma_2 p^*} &=p^* \\
q^*&=\sigma_2 p^*.
\end{split}
\end{equation}
Since $r_1 > \sigma_1r_2$ and $p^* >0 $, we have $p^* > e^{r_1-1-e^{r_1-1}}$.
In view of the assumption, $\sigma_2 e^{r_1-1-e^{r_1-1}}>r_2$ and System (\ref{equilib259}), we have
$
q^* >r_2.
$
Again from (\ref{equilib259}), we have
$
p^* >e^{2r_1-1-e^{r_1-1}}=t_1.
$
Thus (\ref{eqoper2-}) has no other positive equilibrium between $(0,0)$ and $(t_1,r_2)$ and
$$
(0,0) \ll (t_1,r_2) \leq (r_1, r_2) \leq (e^{r_1-1},r_2),
$$
and this verifies (H2) for (\ref{eqoper2}).


We now proceed to verify (H3) for (\ref{eqoper2}). The matrix in (\ref{matrix1}) for (\ref{eqoper2}) is
\begin{equation}\label{matrix17}
\begin{split}
B_{\mu}=(b^{i,j}_{\mu})=\left(
      \begin{array}{ll}
          e^{r_1-\sigma_1r_2} \int_{\mathbb{R}}k_1(s)e^{\mu s}ds& \;\;\;0\\
          r_2 \sigma_2 \int_{\mathbb{R}}k_2(s)e^{\mu s}ds& \;\;\; (1-r_2)\int_{\mathbb{R}}k_2(s)e^{\mu s}ds\\
      \end{array}
    \right)
 \end{split}
\end{equation}
Since $e^{r_1-\sigma_1r_2}>1>1-r_2$, the principal eigenvalue for the matrix is
$$\lambda(\mu)=e^{r_1-\sigma_1r_2} \int_{\mathbb{R}}k_1(s)e^{\mu s}ds$$
and the corresponding positive eigenvector
\begin{equation}\label{egenvectro490}
\vect{\eta}_{\mu}= \left(
                     \begin{array}{c}
                       \nu_{\mu}^{(1)} \\
                       \nu_{\mu}^{(2)} \\
                     \end{array}
                   \right)
=\left(\begin{array}{c}
                                                                 \frac{e^{r_1-\sigma_1r_2}\int_{\mathbb{R}}k_1(s)e^{\mu s}ds-(1-r_2)\int_{\mathbb{R}}k_2(s)e^{\mu s}ds }{r_2 \sigma_2 \int_{\mathbb{R}}k_2(s)e^{\mu s}ds}\\
                                                                                         1
                                                                                                        \end{array}
                                                                                                        \right).
\end{equation}
Because $\int_{\mathbb{R}}k_2(s)e^{\mu s}ds \leq \int_{\mathbb{R}}k_1(s)e^{\mu s}ds$, canceling $\int_{\mathbb{R}}k_2(s)e^{\mu s}ds$ in
$\nu_{\mu}^{(1)}$ leads to
\begin{equation}\label{eq233}
\begin{split}
\nu_{\mu}^{(1)} \geq  \frac{e^{r_1-\sigma_1r_2}-(1-r_2)}{r_2 \sigma_2} \geq \frac{1}{\sigma_2}+\frac{e^{r_1-\sigma_1r_2}-1}{r_2\sigma_2} \geq \frac{1}{\sigma_2}\\
\end{split}
\end{equation}
It is clear now that (H3)(i) holds.  We can proceed to  verify (H3)(ii) for (\ref{eqoper2+}).  Let
$$(p,q)=(\min\{e^{r_1-1}, \nu_{\mu}^{(1)} \alpha \},\min\{r_2, \alpha\}), \alpha >0.$$
Since $e^{q-\sigma_2 p} \geq  1+ q-\sigma_2 p$, we need to show that
\begin{equation}\label{eqg_13}
\begin{split}
h^+(p)e^{r_1-\sigma_1 r_2+\sigma_1 q} & \leq e^{r_1-\sigma_1r_2} p\\
r_2-\big(r_2-q\big)e^{q-\sigma_2 p} & \leq r_2 \sigma_2 p + (1-r_2)q + q(q-\sigma_2p)\\
& \leq  r_2 \sigma_2 p + (1-r_2)q
\end{split}
\end{equation}
Therefore, it is easy to see that we only need to verify that
\begin{equation}\label{eqg_13-1}
\begin{split}
q & \leq  \sigma_2 p\\
\end{split}
\end{equation}
and
\begin{equation}\label{eqg_13-2}
\begin{split}
\frac{h^{+}(p)}{p} & \leq e^{-\sigma_1 q}
\end{split}
\end{equation}
For (\ref{eqg_13-1}), we need to consider the two cases: $p=e^{r_1-1}$ and $p=\nu_{\mu}^{(1)} \alpha$. If $p=e^{r_1-1}$, then
\begin{equation}\label{eq787}
q \leq r_2 \leq \sigma_2 e^{r_1-1}
\end{equation}
which is true by the assumption ($r_2< \sigma_2 e^{r_1-1-e^{r_1-1}}$). If $p=\nu_{\mu}^{(1)} \alpha$, then
$
q \leq \alpha \leq \sigma_2 \nu_{\mu}^{(1)} \alpha,
$
which is true because of (\ref{eq233}).

In order to verify (\ref{eqg_13-2}), first assume that $p \in (0,1)$, then $h^{+}(p)=pe^{-p}$ and $p=\nu_{\mu}^{(1)} \alpha$ since $e^{r_1-1}>1$.
Since $e^{-\sigma_1 \alpha } \leq  e^{-\sigma_1 q}$, it suffices to verify that
$
e^{-\nu_{\mu}^{(1)} \alpha} \leq e^{-\sigma_1 \alpha },
$
 which is true because of (\ref{eq233}) and $\sigma_2\sigma_1<1.$ For the case $p \geq 1$ we have $h^{+}(p)=e^{-1}$.
Again since
\begin{equation}\label{eq8989}
e^{- r_2 } \leq e^{-\sigma_1 r_2 } \leq  e^{-\sigma_1 q},
\end{equation}
it suffices to verify
$
\frac{e^{-1}}{p} \leq e^{-1} \leq e^{-r_2 },
$
 which holds because $r_2<1$.

It remains to verify (H3)(ii) for (\ref{eqoper2-}).  Let
$$(p,q)=(\min\{t_1, \nu_{\mu}^{(1)} \alpha \},\min\{r_2, \alpha\}), \alpha >0.$$

For (\ref{eqg_13-1}), we need to consider the two cases: $p=t_1$ and $p=\nu_{\mu}^{(1)} \alpha$. If $p=t_1$, from the assumptions, we have
\begin{equation}\label{eq789}
q \leq r_2 \leq \sigma_2  e^{r_1-1 -e^{r_1-1}} < \sigma_2  e^{2 r_1-1 -e^{r_1-1}} = \sigma_2 t_1= \sigma_2 p
\end{equation}
and then (\ref{eqg_13-1}) holds. If $p=\nu_{\mu}^{(1)} \alpha$, then
$
q \leq \alpha \leq \sigma_2 \nu_{\mu}^{(1)} \alpha,
$
which is true because of (\ref{eq233}).

We must verify (\ref{eqg_13-2}) (with $h^{+}$ being replaced by $h^{-}$) for (\ref{eqoper2-}).  If $0<p <t_0$,
then $h^{-}(p)=pe^{-p}$ and $p=\nu_{\mu}^{(1)} \alpha$ because of (\ref{eq9989}).
Since $e^{-\sigma_1 \alpha } \leq  e^{-\sigma_1 q}$, it suffices to verify that
$
e^{-\nu_{\mu}^{(1)} \alpha} \leq e^{-\sigma_1 \alpha },
$
which is true because of (\ref{eq233}) and $\sigma_2\sigma_1<1.$ For the case that $p \geq t_0$,
then $h^{-}(p)=h(t_0)$. From the definition of $h^-$ and (\ref{eq8989}) we see that it suffices to verify that
\begin{equation}\label{eq1000}
\frac{h^{-}(p)}{p} \leq \frac{h(t_0)}{t_0} = e^{-t_0} \leq e^{-\sigma_1 r_2 }
\end{equation}
holds, which follows from Expression (\ref{eq9989}) and the assumption,
\begin{equation}\label{eq1001}
 e^{-t_0} \leq e^{-e^{r_1-1-e^{r_1-1}}} \leq e^{-\sigma_1 r_2 }.
\end{equation}
We observe here that the assumption $\sigma_1 r_2< e^{r_1-1-e^{r_1-1}}$ can be relaxed as long as (\ref{eq1000})($t_0 \geq \sigma_1 r_2$) holds.

To verify (H3)(iii), we note that $h^-(p)=h^+(p)=h(p)$  for $p$ small, and conclude from Lemma \ref{estimageg}, for sufficiently larger $k$, there is a small $\omega  \gg 0 $, if  $0 \leq (p,q) \leq \omega$,
$
f(p,q) \geq f(p,0) \geq (1-\frac{1}{k})e^{r_1-\sigma_1r_2} p
$
and
$
g(p,q) \geq (1-\frac{1}{k})r_2 \sigma_2  p + (1-\frac{1}{k})(1-r_2) q.
$

This concludes the proof of Theorem \ref{th33} since the conditions (H1-H3) have been verified.
\epf

\end{document}